\documentclass{ws-procs975x65}

\begin{document}

\title{RENORMALIZATION GROUP TRAJECTORIES BETWEEN TWO FIXED POINTS}

\author{ABDELMALEK ABDESSELAM}

\address{Department of Mathematics, University of Virginia,\\
P. O. Box 400137, Charlottesville, VA 22904-4137, USA\\
E-mail: malek@virginia.edu}

\begin{abstract}
We report on our recent rigorous construction of complete renormalization group
trajectories between two fixed points for the three-dimensional phi-four model with modified propagator
considered by Brydges, Mitter and Scoppola (BMS). These are discrete critical trajectories which
connect the ultraviolet Gaussian fixed point to the nontrivial BMS infrared fixed point which
is an analogue of the Wilson--Fisher fixed point. The renormalization group map is defined rigorously
and nonperturbatively, without using the hierarchical approximation. The trajectories are constructed by a
fixed point argument in a  suitable Banach space of sequences, where one perturbs a nonlinear one-dimensional iteration.
\end{abstract}

\keywords{renormalization group, crossover, heterclinic orbit, invariant manifolds, massless flow,
Wilson-Fisher fixed point.}

\bodymatter

\section{Global Dynamics of Wilson's Renormalization Group}
A central mathematical problem posed by quantum field theory (QFT) is that of giving
a rigorous meaning to functional integrals such as
 \[
\int_{\mathcal{F}}\ D
\phi \cdots
e^{-\int_{\mathbb{R}^d} [\frac{1}{2}(\nabla\phi)^2(x)+\mu \phi(x)^2+g \phi(x)^4]dx}\ .
\]
In this example corresponding to the so-called $\phi^4$ theory,
$\mathcal{F}$ is the infinite-dimensional space of functions form $\mathbb{R}^d$ to $\mathbb{R}$,
$D\phi$ is the Lebesgue measure on $\mathcal{F}$, and the dots stand for some observable or functional of the
field $\phi$. A natural approach to this problem is to try to construct such a measure on functions or rather
distributions in $\mathcal{S}'(\mathbb{R}^d)$ 
by a scaling limit of lattice theories on $(a\mathbb{Z})^d\subset \mathbb{R}^d$.
This corresponds to imposing a cut-off $\frac{1}{a}$ on momenta in Fourier space.
One can easily
rescale to a unit lattice and see the
approximants to the continuum theory as living on the lattice $\mathbb{Z}^d$ instead of $(a\mathbb{Z})^d$.
These approximants become points in the space of all possible unit cut-off theories. In essence, Wilson's renormalization group
(RG) is a dynamical system acting on this space.
If 
$d\nu$ is a measure on random fields $\phi$ with $\hat{\phi}(p)=0$ if $|p|>1$, i.e. with sharp
unit cut-off on momenta, one
introduces a magnification ratio $L>1$
and
splits the field as $\phi=\zeta+\phi_{\rm low}$.
The fluctuation field
$\zeta$ corresponds to keeping Fourier modes with $L^{-1}<|p|\le 1$,
whereas the low momentum or background field
$\phi_{\rm low}$ has Fourier modes with $|p|\le L^{-1}$.
One can then
integrate over $\zeta$, and obtain the marginal probability distribution on $\phi_{\rm low}$
coming from the original probability distribution $d\nu$.
Finally one rescales back to unit lattice by letting $\psi(x)=L^{[\phi]} \phi_{\rm low}(Lx)$,
and considering the resulting probability distribution $d\nu'$ on $\psi$.
The
RG map is the transformation {$d\nu\longrightarrow d\nu'$}.
Important features of this dynamical system are the notions of
fixed points, eigenvalues of linearized RG around them, local stable and unstable manifolds etc.
They play a key role in controlling scaling limits and constructing continuum functional integrals
by lattice approximations.
The features just listed which have also been the most studied ones are local, i.e., pertaining to the vicinity
of single fixed points. Much less is known about
global features of the RG dynamical system, e.g., heteroclinic trajectories between fixed points.
This is the main object of this article which briefly reports on recent progress we made in this area~\cite{AbdCMP07}.

\section{Rigorous Results (Selection)}
There is a long history of rigorous results on the RG dynamical system.
We will only mention a selection of such results which gives the best idea of
the context into which the results of our article~\cite{AbdCMP07} fit.

The RG exponents for the Gaussian fixed point, the existence of a nontrivial infrared fixed point in `$4-\epsilon$' dimensions, as well as the construction
of its local stable manifold, were obtained by Bleher and Sinai~\cite{BlS1,BlS2},
in the hierchical model (HM)
approximation.
Similar constructions were also provided by
Collet and Eckmann~\cite{ColEck1,ColEck2}, Gaw\c{e}dzki and Kupiainen~\cite{GKcmp83,GKjsp84}.
Furthermore, the two-dimensional local unstable manifold of the Gaussian fixed point
for the $\phi^4_3$ HM was constructed~\cite{Pereira}.
The hierarchical $\phi^4_3$ infrared fixed point (i.e. for $\epsilon=1$)
was also constructed~\cite{KochW}.
The appearance of new fixed points at dimensions $d=2+\frac{2}{n-1}$, $n=3,4,\ldots$ in the local
potential approximation was established by Felder~\cite{Felder}.

As per the Euclidean model, i.e., the full model without hierarchical or local potential aproximations,
the existence of a nontrivial infrared fixed point as well as the control of its local stable manifold
was provided by Brydges, Dimock and Hurd~\cite{BDHfp}.
Later, a similar result for a nicer `$(4-\epsilon)$-dimensional' model was derived by Brydges, Mitter
and Scoppola~\cite{BMS}.

The previous results concern local features of the RG. Rigorous results of a global nature are more
scarce.
One can mention the proof of
uniqueness of the infrared fixed point in the local potential approximation in dimension
$3\le d<4$ by Lima~\cite{Lima}.
Closest to our result is the construction of 
the massless Gross-Neveu model in `$2+\epsilon$' dimension by Gaw\c{e}dski and Kupiainen~\cite{GKnonren}.
Their result amounts to controlling a heteroclinic trajectory joining a nontrivial ultraviolet
fixed point to a trivial infrared fixed point, for a Fermionic QFT.
In our article~\cite{AbdCMP07} which concerns the Bosonic BMS model~\cite{BMS}, we 
obtained the construction of discrete heteroclinic trajectories joining a Gaussian ultraviolet fixed point
to a nontrivial infrared fixed point.
One should finally also mention related work on massless nontrivial QFT models
such as $\phi^4$ in two and three dimensions~\cite{McBryanR,GJcrit,Destruct}
or the Thirring model~\cite{BFMastro}.

\section{The BMS Model}
The BMS model for a scalar field $\phi:\mathbb{R}^3\longrightarrow \mathbb{R}$
corresponds to the functional integral
\[
Z=\underbrace{\int D\phi\ e^{-\frac{1}{2}\langle\phi, (-\Delta)^{\frac{3+\epsilon}{4}}
\phi\rangle_{L^2(\mathbb{R}^3)}}}_{\rm Gaussian\ measure}
{\ }^{-}  \overbrace{{\ }^{\int dx (g :\phi^4(x):+\mu :\phi^2(x):)}}^{{\rm potential}\ V(\phi)}\ .
\]
The free propagator is $(-\Delta)^{-\frac{3+\epsilon}{4}} (x,y)\sim \frac{1}{|x-y|^{2[\phi]}}$
where $[\phi]=\frac{3-\epsilon}{4}$ denotes the canonical scaling dimension of the field.
One also has a multiscale decomposition of this propagator, in the form
$\int_0^{\infty} \frac{dl}{l} l^{-2[\phi]} \ u\left(\frac{x-y}{l}\right)$
with $u$ a smooth, compactly supported, rotationally symmetric function which is
nonnegative in both $x$ and $p$.
Instead of a lattice regularization, it is more convenient to use continuous cut-offs, by
taking $C(x-y)=\int_1^{\infty} \frac{dl}{l} l^{-2[\phi]} \ u\left(\frac{x-y}{l}\right)$
as a free propagator.
The high versus low momentum 
split corresponds to the decomposition $C(x-y)=\Gamma(x-y)+C_{L^{-1}}(x-y)$
with $C_{L^{-1}}(x-y)=L^{-2[\phi]}C(L^{-1}(x-y))$. The fluctuation covariance therefore
is $\Gamma(x-y)=\int_1^{L} \frac{dl}{l} l^{-2[\phi]} \ u\left(\frac{x-y}{l}\right)$.
One can rewrite accordingly the original Gaussian measure as a convolution
$d\mu_C=d\mu_\Gamma \star d\mu_{C_{L^{-1}}}$. 
The functional integral, with integrand $\mathcal{Z}(\phi)$, becomes
\[
Z=\int d\mu_C(\phi)\ \mathcal{Z}(\phi)=
\int d\mu_{C_{L^{-1}}}(\psi) d\mu_{\Gamma}(\zeta)\ \mathcal{Z}(\psi+\zeta)
=\int d\mu_C (\phi)\ (\mathcal{R}\mathcal{Z})(\phi)
\]
where
$(\mathcal{R}\mathcal{Z})(\phi)=\int d\mu_{\Gamma}(\zeta)\ \mathcal{Z}(\phi_{L^{-1}}+\zeta)$
and
$\phi_{L^{-1}}(x)=L^{-[\phi]}\phi(L^{-1}x)$.
Now the
RG map can be seen as acting on integrands: $\mathcal{Z}\longrightarrow \mathcal{R}\mathcal{Z}$.

\section{Good Infinite Volume Coordinates}
An important ingredient for the rigorous implementation of the RG is a good set of coordinates
for the dynamical variable $\mathcal{Z}$, which behave well in the infinite volume limit.
Such a formalism was introduced by Brydges and Yau~\cite{BY}.
One uses the decomposition of $\mathbb{R}^3$ into unit cubes or cells defined by a lattice $\mathbb{Z}^3$.
If $\Lambda$ is the large finite box where the model is defined, one writes the following
polymer representation for the integrand:
\[
\mathcal{Z}(\Lambda,\phi)=
\sum_{n=0}^\infty
\frac{1}{n!} \sum_{{X_1,\ldots,X_n}\atop{{\rm disjoint\ in}\ \Lambda}}
\exp\left[
-\int_{\Lambda\backslash(\cup X_i)} dx \{g :\phi^4(x):_C+ \mu :\phi^2(x):_C\}
\right]
\]
\begin{equation}
\times K(X_1,\phi|_{X_1})\cdots K(X_n,\phi|_{X_n})\ .
\label{polyrep}
\end{equation}
The $X_i$ are polymers, i.e., connected finite unions of closed unit cells.
The functionals $K(X,\phi)$ are local in the sense that they only depend
on the restriction of the field to the sets $X$.
The coordinates for $\mathcal{Z}$ now are triples $(g,\mu,K)$
where $g,\mu$ are the running couplings for the $\phi^4$ and mass terms, and
$K=(K(X,\cdot))_{X\ {\rm polymer}}$
is a collection of local functionals.
The need to explicitly extract second order perturbation theory, in order to analyse the RG
dynamics, imposes an additional rewriting of the form
\[
K(X,\phi)=g^2 [{\rm explicit\ complicated\ formula}] e^{-V(X,\phi)}
+R(X,\phi)\ .
\]
So the dynamical variable becomes the collection of functionals $R$
which can be thought of as an $O(g^3)$ remainder.
As a result, the RG for the BMS model can be written as an explicit (albeit complicated)
map $(g,\mu,R)\longrightarrow (g',\mu',R')$.
It has the form
\begin{eqnarray*}
g' & = & L^\epsilon g-L^{2\epsilon}a(L,\epsilon) g^2+\xi_g(g,\mu,R) ,\\
\mu' & = & L^{\frac{3+\epsilon}{2}}\mu+\xi_\mu(g,\mu,R)\ ,\\
R' & = & \mathcal{L}^{(g,\mu)}(R)+\xi_R(g,\mu,R)\ .
\end{eqnarray*}
The quantity $a(L,\epsilon)$ is $\sim \log L$. The $\xi$'s are small nonlinear remainder
terms. The $\mathcal{L}^{g,\mu}$ stands for a $(g,\mu)$-dependent linear operator in $R$.
Its crucial property is that it is contractive, which follows from the so-called ``extraction step''
in the definition of the RG map. This is the analogue in the present setting of the traditional
BPHZ subtractions in renormalization theory~\cite{Rivass}.
It removes the dangerous local part of the two and four point dependence in the low momentum field
$\phi_{L^{-1}}$, from the functionals $R$ and transfers them to the exponentiated potential. 
This relies on the nonuniqueness of the polymer representation of
Eq.~\ref{polyrep} for a given integrand
$\mathcal{Z}$.

Ignoring the $\xi$ remainders makes for a simplified RG map with easy dynamics.
In addition to the trivial fixed point $(0,0,0)$, it has a fixed point at $(\bar{g}_\ast,0,0)$
with
$\bar{g}_\ast=\frac{L^\epsilon -1}{L^{2\epsilon}a}\sim \epsilon$.
It also has heteroclinic trajectories $(\bar{g}_n,0,0)_{n\in\mathbb{Z}}$
parametrized by the single number $\bar{g}_0$ in the interval $(0,\bar{g}_\ast)$,
and constructed using the one-dimensional iteration by the function $f(x)=L^\epsilon x-L^{2\epsilon} a x^2$.
The BMS fixed point was obtained as a perturbation of the approximate fixed point $(\bar{g}_\ast,0,0)$.
Similarly, the true RG connecting orbits are constructed by perturbing the above approximate trajectories.  
Our main result~\cite{AbdCMP07} is as follows.

\begin{theorem}
If the bifurcation parameter $\epsilon>0$ is small enough, for any
$\omega_0\in]0,\frac{1}{2}[$, there exists a (locally unique) complete
trajectory $(g_n,\mu_n,R_n)_{n\in\mathbb{Z}}$
for the RG map
such that
$\lim\limits_{n\rightarrow -\infty}
(g_n,\mu_n,R_n)=(0,0,0)$
the Gaussian ultraviolet fixed point, and
$\lim\limits_{n\rightarrow +\infty}
(g_n,\mu_n,R_n)=(g_\ast,\mu_\ast,R_\ast)$
the BMS nontrivial infrared fixed point,
and
determined by the `initial condition' at unit scale
$g_0=\omega_0\bar{g}_\ast$.
\end{theorem}

\section{Idea of The Proof}
One writes $(g_n,\mu_n,R_n)=(\bar{g}_n+\delta g_n,\mu_n,R_n)$, so the trajectory
equations are rephrased in terms of the deviation variables $(\delta g_n,\mu_n,R_n)$
relating the true trajectory to the approximate one.
One has to enforce three boundary conditions:
1) $\mu_n$ does not blow up when $n\rightarrow +\infty$ (infrared),
2) $R_n$ does not blow up when $n\rightarrow -\infty$ (ultraviolet),
3) $\delta g_0=0$ (unit scale).
One iterates the flow equations either backwards or forwards towards the appropriate boundary condition.
For instance, the $\mu_n$ term is reexpressed by forward iteration as:
\begin{equation}
\mu_n= -\sum\limits_{p\ge n}
L^{-\left(\frac{3+\epsilon}{2}\right)(p-n+1)}
\ \xi_{\mu}(\bar{g}_p+\delta g_p,\mu_p,R_p)
\label{mueq}
\end{equation}
The $R$ is iterated backwards. Finally the $g_n$ is iterated backwards if $n>0$, and forwards if $n<0$.
The four `integral' equations, including Eq.~\ref{mueq}, obtained in this way are interpreted as a fixed point equation
in a big Banach space of two-sided sequences. The sought for trajectories are thus obtained using
Banach's contraction mapping theorem.

\section{Functional Analysis, Norms}
Let $\Delta$ denote a (closed) unit cell. There is a Sobolev imbedding $W^{4,2}(\stackrel{\circ}{\Delta})\hookrightarrow C^2(\Delta)$.
The fields $\phi$ live in 
Hilbert spaces ${\rm Fld}(X)=\bigoplus_{\Delta\subset X}W^{4,2}(\stackrel{\circ}{\Delta})$
where  extra $C^2$ gluing conditions are imposed, using the Sobolev imbedding.
The norm on a field is given by $||\phi||_{{\rm Fld}(X)}^2
=\sum_{\Delta\subset X}
\sum_{|\nu|\le 4}
||\partial^\nu \phi_\Delta||_{L^2(\stackrel{\circ}{\Delta})}^2$.
The fluctuation measure $d \mu_\Gamma$ is realized in the Hilbert spaces ${\rm Fld}(X)$.
One also needs a second field norm $||\phi||_{C^2(X)}
=\sup\limits_{x\in X}
\max\limits_{|\nu|\le 2}
|\partial^\nu \phi(x)|$.
While the
$\phi$'s are real-valued, the
functionals can be complex valued and are
compared using norms of the form
\[
||K||=\sup_{\Delta_0}
\sum_{X\supset \Delta_0}
L^{5|X|} \sup_{\phi\in{\rm Fld}(X)}
\left\{
e^{-\kappa \sum_{\Delta\subset X}\sum_{1\le |\nu|\le 4}
||\partial^\nu \phi||_{L^2(\stackrel{\circ}{\Delta})}^2}
\right.
\]
\[
\times
\left.
\sum_{0\le n\le 9} \frac{(c {g^{-\frac{1}{4}}})^n}{n!}
\sup_{\phi_1,\ldots,\phi_n\in{\rm Fld}(X)\backslash\{0\}}
\frac{|D^n K(X,\phi;\phi_1,\ldots,\phi_n)|}{||\phi_1||_{C^2(\Delta)}\cdots
||\phi_n||_{C^2(\Delta)}}
\right\}
\]
where $D^n K$ denotes the $n$-th order differential.
One of the main technical difficulties, or fibered norm problem, is that these
norms depend on the dynamical variable $g$.
The way around it is to use
the approximate solution $\bar{g}_n$ to calibrate them~\cite{AbdCMP07}.

\section{Perspectives}
Many open problems in the continuation of this work remain.
One should study more refined dynamical systems features of the BMS model such as the construction of the full heteroclinic invariant curve. One should study its
regularity properties and
smoothness at $g=0$ (asked by K. Gaw\c{e}dzki).
There is the problem of constructing the
correlation functions for the primary field $\phi$
and also for
composite fields and study possible anomalous dimensions (there are preliminary results on this by P. K. Mitter).
A good testing ground for these more ambitious goals is the case of a hierarchical version of the BMS model (ongoing
work by the author together with G. Guadagni and Ph.~D. student A. Chandra).
This could lead to the construction of the massless Euclidean BMS model over the $p$-adics.
If the previous hurdles are cleared, one should also study the analytic continuation to Minkowski
space (in the Archimedean case).
Another interesting project would
be to provide an alternate construction of the critical BMS theories
using recent phase space expansion techniques~\cite{Abdthese}. Finally the problem of controlling
complete RG trajectories (from the UV to the IR end of the scale spectrum), here addressing
the BMS model, is a worthy and certainly difficult task for many more QFT models:
Gross-Neveu in $2d$,
$\phi_3^4$ at large $N$,
noncommutative vulcanized
$\phi_4^4$,
the
$2d$ $\sigma$-model,
Yang-Mills in $4d$.
The inspiring talk by S. Weinberg on QFT and the asymptotic safety scenario,
at this congress, also leads one to wonder if one could devise
a rigorous RG framework in order to study
a nontrivial ultraviolet fixed point in quantum gravity\ldots

\section*{Acknowledgments}
The research presented in this article is supported by the National Science Foundation
under grant \# DMS--0907198.


\begin{thebibliography}{9}
\bibitem{AbdCMP07} A. Abdesselam, {\em Commun. Math. Phys.} {\bf 276},
                   727 (2007).
\bibitem{BlS1}     P.~M. Bleher and Ja.~G. Sinai, {\em Commun. Math. Phys.}
                   {\bf 33}, 23 (1973).
\bibitem{BlS2}     P.~M. Bleher and Ya.~G. Sinai, {\em Commun. Math. Phys.}                   
                    {\bf 45}, 247 (1975).
\bibitem{ColEck1}  P. Collet and J.-P. Eckmann, {\em Commun. Math. Phys.}                  
                   {\bf 55}, 67 (1977).
\bibitem{ColEck2}  P. Collet and J.-P. Eckmann, {\it A Renormalization Group Analysis
                   of The Hierarchical Model in Statistical Mechanics}, Lect. Notes in Phys.
                   {\bf 74} (Springer, Berlin--New York, 1978).
\bibitem{GKcmp83}  K. Gaw\c{e}dzki and A. Kupiainen, {\em Commun. Math. Phys.}                                   
                   {\bf 89}, 191 (1983).
\bibitem{GKjsp84}  K. Gaw\c{e}dzki and A. Kupiainen, {\em J. Stat. Phys.}                 
                   {\bf 35}, 267 (1984).
\bibitem{Pereira}  E.~A. Pereira, {\em J. Math. Phys.} {\bf 34}, 5770 (1993).                   
\bibitem{KochW}    H. Koch and P. Wittwer, {\em Commun. Math. Phys.}                   
                   {\bf 106}, 495 (1986).
\bibitem{Felder}   G.~Felder, {\em Commun. Math. Phys.} {\bf 111}, 101 (1987).
\bibitem{BDHfp}    D.~C. Brydges, J. Dimock and T.~R. Hurd, {\em Commun. Math. Phys.}
                   {\bf 198}, 111 (1998).
\bibitem{BMS}      D.~C. Brydges, P.~K. Mitter and B. Scoppola, {\em Commun. Math. Phys.}
                   {\bf 240}, 281 (2003).
\bibitem{Lima}     P.~C. Lima, {\em Commun. Math. Phys.} {\bf 170}, 529 (1995).
\bibitem{GKnonren} K. Gaw\c{e}dski and A. Kupiainen, {\em Nuclear Phys. B} {\bf 262},
                   33 (1985). 
\bibitem{McBryanR} O.~A. McBryan and J. Rosen, {\em Commun. Math. Phys.} {\bf 51},
                   97 (1976).
\bibitem{GJcrit}   J. Glimm and A. Jaffe, {\em Commun. Math. Phys.} {\bf 52}, 203 (1977).
\bibitem{Destruct} D.~C. Brydges, J. Fr{\"o}hlich and A.~D. Sokal,
                   {\em Commun. Math. Phys.} {\bf 91}, 141 (1983).
\bibitem{BFMastro} G. Benfatto, P. Falco and V. Mastropietro, {\em Commun. Math. Phys.}
                   {\bf 273}, 67 (2007).
\bibitem{BY}       D.~C. Brydges and H.~T. Yau, {\em Commun. Math. Phys.} {\bf 129}, 351 (1990).                  
\bibitem{Rivass}   V. Rivasseau, {\it From Perturbative to Constructive
                   Renormalization} (Princeton University Press, Princeton NJ, 1991).
\bibitem{Abdthese} A. Abdesselam, {\it Renormalisation Constructive Explicite}, Ph. D. thesis
                   (Ecole Polytechnique, Palaiseau, 1997).
                                      
\end{thebibliography}
\end{document}